# Acoustic gap placement in second-language read speech production

Peyman Jahanbin
Department of Curriculum and Teaching, University of Kansas, Lawrence, Kansas, United States of America
Corresponding author: peyman@ku.edu
ORCID: 0000-0002-5692-3379
Short title: Acoustic gap placement in second-language read speech

## Abstract

Speakers can differ not only in how much silence they produce but also in where interruptions fall, information that global pause counts obscure. We analyzed 115 publicly available Speech Accent Archive recordings of the same read passage: 57 speakers whose archive metadata listed Mandarin as their first language and a mainland-China birthplace, and 58 American English first-language speakers born in the U.S. Midwest. Word-level forced alignment was used to extract interword acoustic gaps and classify each position as punctuation-marked or unpunctuated. At a 250 ms threshold, punctuation-marked gaps were similar for the Mandarin and English groups (means 5.56 and 5.47), whereas unpunctuated-position gaps were more frequent in the Mandarin group (means 3.67 and 0.43; adjusted rate ratio 11.98, 95% confidence interval 6.53–22.00). A crossed-effects logistic model with random intercepts for speakers and passage positions confirmed that the contrast was disproportionately concentrated at unpunctuated positions (interaction odds ratio 9.44, 95% confidence interval 5.23–17.02). The pattern persisted at a 500 ms threshold, after removing positions adjacent to punctuation, after excluding severe transcript-deviation cases, and when analysis was restricted to acoustically confirmed silence. Nearly half of the Mandarin group's unpunctuated gaps occurred at positions classified as within-phrase in an exploratory single-coder annotation. In this fixed-passage corpus, acoustic gap placement relative to textual and syntactic structure distinguished the groups more clearly than punctuation-marked pausing. The results support placement-sensitive measurement as a reproducible dimension of breakdown fluency while not establishing a causal mechanism, proficiency difference, or perceptual consequence.

Keywords: second-language speech; breakdown fluency; pause placement; acoustic gaps; forced alignment; read speech

## Introduction

Oral fluency is a multidimensional property of spoken performance rather than a single rate or score [1]. Speech unfolds over time, and temporal measures describe complementary aspects of that performance. Yet the measures most often reported at the speaker level, such as total pause count, mean pause duration, or speech rate, summarize across an entire sample of speech. They can establish that two performances differ in the quantity or duration of interruption, but they do not reveal whether those interruptions occupy comparable locations in the linguistic material.

The present study concerns utterance fluency, not cognitive fluency or perceived fluency. The distinction among these constructs [2] is consequential: utterance fluency consists of observable properties of a speech signal, cognitive fluency concerns the efficiency of the processes that support production, and perceived fluency is a listener judgment. Within utterance fluency, speed measures capture how quickly speech is delivered, breakdown measures capture interruptions, and repair measures capture reformulations and false starts [3,4]. Pause frequency, pause duration, speech rate, and repair make different contributions to listeners' judgments and should not be treated as interchangeable indicators [5–8]. Because the Speech Accent Archive provides recordings but no experimental processing measures or listener ratings, the claims here are deliberately restricted to breakdown fluency in the acoustic record.

Placement is an important but easily discarded property of breakdown fluency. Silence in speech is not randomly distributed: pauses frequently coincide with structural boundaries and can contribute to the organization of an utterance [9,10]. Consequently, two speakers with the same number of gaps can present different timing profiles. One may place most gaps at sentence boundaries or punctuation marks, whereas another may interrupt word sequences that contain no visible boundary cue. A global count assigns these performances the same value even though the interruptions are distributed differently across the input.

Prior second-language research provides empirical reasons to preserve this information. Mid-clause or within-clause pausing has differentiated learner and first-language speech [11,12], and experimental work shows that listeners are sensitive to pause location, not only pause frequency or duration [13]. A Mandarin–English read-speech comparison likewise related pausing to boundary level and syntactic complexity [14]. De Jong found that first- and second-language speakers differed within utterances but not at utterance boundaries, demonstrating that a group contrast can depend on where a pause occurs [15]. More recent work has proposed continuous syntactic-distance measures of pause location [16] and has examined location-sensitive pausing in relation to comprehensibility [17]. Together, this literature suggests that the distribution of gaps relative to linguistic structure is a dimension of performance that global totals can obscure.

Read speech offers a particularly transparent setting in which to study placement. Lexical content, word order, and visible punctuation are held constant, and each interword position can be compared directly across speakers. Punctuation and textual organization exert strong effects on pause placement in reading [18], although punctuation is neither a complete syntactic representation nor a transcription of prosodic phrasing. In spontaneous speech, content selection, formulation, lexical retrieval, repair, and interaction vary together [19,20]. A fixed passage removes much of that between-speaker variation and has an established role in objective temporal assessment [21]. The gain in control is accompanied by a clear limit: a pattern observed in one read passage cannot be assumed to generalize to spontaneous speech or to other texts.

The closest comparison to the present study is Alfaifi and colleagues' analysis of 10 Mandarin–English late bilinguals and 10 monolingual English speakers reading "The North Wind and the Sun" [14]. That study examined pause duration in relation to syntactic complexity and selected punctuation or prosodic boundaries. It showed that boundary level and syntactic complexity were relevant to pausing in this population. The present study asks a complementary full-passage question. Instead of beginning with selected likely pause sites, it extracts every positive interword interval and contrasts all punctuation-marked boundaries with every unpunctuated position in the passage. This design can show whether a group difference is broadly distributed, concentrated at visible boundaries, or concentrated away from them.

Any pause analysis also depends on the duration threshold used to distinguish a counted interruption from ordinary segmental timing. A 250 ms cutoff is well established in second-language fluency research [22], while recent evidence indicates that the threshold most predictive of proficiency or perceived fluency varies across speaking contexts and outcomes [23]. The present analysis therefore uses 250 ms as its primary threshold and 500 ms as a deliberately stricter sensitivity threshold. Because forced-alignment intervals are not automatically equivalent to silent pauses, the analysis also evaluates the acoustic character of the aligner-defined intervals using an additional signal-based criterion.

Against this background, the study addresses four questions: whether the two groups differ in total interword acoustic-gap count and duration; whether any group difference depends on whether the position is punctuation-marked or unpunctuated; whether the unpunctuated-position contrast survives threshold, adjacency, transcript, manual-inspection, and acoustic-confirmation checks; and whether it extends to positions classified as within-phrase. The directional expectation was that the Mandarin–English contrast would be larger at unpunctuated than at punctuation-marked positions. The syntactic refinement was treated as exploratory, and no gap location was interpreted as direct evidence of a particular cognitive mechanism.

## Materials and methods

### Corpus, participant selection, and sample size

Recordings came from the Speech Accent Archive [24], a publicly accessible collection in which speakers read the same elicitation paragraph. The canonical “Please call Stella” passage used here contains 69 words. Earlier work used recordings from the same archive to compare Mandarin and American English speakers through MFCC-based modeling [25], with a companion preprint documenting the reproducible analytical pipeline [26]. The project corpus was assembled before the present placement analysis, and participant eligibility was defined from the archive’s own biographical fields rather than from any acoustic outcome.

The Mandarin-group rule required the archive page to identify Mandarin as the native language and to list a birthplace in mainland China. A record listing Taiwan, Hong Kong, Macao, Singapore, or another country would not satisfy this geographic rule. The English-group rule required English as the native language and a U.S. birthplace in the Census Midwest region: Illinois, Indiana, Iowa, Kansas, Michigan, Minnesota, Missouri, Nebraska, North Dakota, Ohio, South Dakota, or Wisconsin. The project corpus was assembled in March 2026 and included every Speech Accent

Archive recording available at that time that met the stated native-language and birthplace criteria. Because the archive is continually updated, it does not include recordings added afterward; the July 15, 2026 audit re-verified every retained record against the stated criteria.

On July 15, 2026, we rechecked all 118 project participant codes against their individual archive pages. The audit recorded the public speaker code, native language, birthplace, age, sex, other available language-history fields, selection decision, retrieval date, and participant-page URL. Three Mandarin-labeled project records did not meet the intended mainland-China birthplace rule: mandarin11 was born in Taiwan, mandarin12 in Singapore, and mandarin20 in Taiwan. They were excluded before any model was fit. The resulting analytic sample contained 115 speakers: 57 Mandarin first-language speakers born in mainland China and 58 American English first-language speakers born in the Midwest. No recording was included or excluded because of its number, duration, or location of acoustic gaps. S2 Data supplies the complete audit, including the three exclusions and source URLs.

All 115 retained recordings had usable audio and complete 69-word forced alignments. No a priori power analysis determined the sample size; the analysis used the eligible recordings in the established project corpus after the geographic correction. The emphasis is therefore on estimated effect sizes, uncertainty intervals, and convergence across sensitivity analyses rather than on post hoc power. Age and sex were transcribed from the archive and included as covariates because age differed descriptively between groups and both variables were available for every speaker. Other archive fields, including age of English onset, learning method, residence history, and additional languages, are self-reported, heterogeneous, and not sufficient to represent proficiency or language experience as standardized covariates. The adjusted coefficients should consequently be interpreted as age- and sex-adjusted associations, not as causal group effects.

**Table 1. Participant and corpus summary**

| Group | n | Age, mean (SD), range | Female | Male | Duration, mean (SD), s | Nominal passage rate, mean (SD), words/s |
|---|---|---|---|---|---|---|
| American English L1 | 58 | 35.65 (15.96), 18–75 | 32 | 26 | 22.97 (3.64) | 3.07 (0.41) |

| Group | n | Age, mean (SD), range | Female | Male | Duration, mean (SD), s | Nominal passage rate, mean (SD), words/s |
|---|---|---|---|---|---|---|
| Mandarin L1, mainland China | 57 | 29.16 (7.82), 18–49 | 35 | 22 | 31.52 (7.63) | 2.31 (0.55) |

*Note.* Duration is total recording duration. Nominal rate is 69 divided by total duration and is descriptive context, not articulation rate. Age and sex are archive-recorded metadata fields.

**Ethics and privacy**

This retrospective observational study used only pre-existing recordings and biographical fields that the Speech Accent Archive makes publicly accessible for research. The study involved no interaction or intervention with individuals and no access to private information; institutional ethics review and additional informed consent for this secondary analysis were therefore not required. All retained speakers were adults (ages 18–75) according to archive metadata. No individuals were contacted, and the study did not attempt to recover names or infer information beyond the public archive record. The article does not redistribute audio. Its supporting files contain public archive codes and source URLs together with derived timing variables, model inputs, and analysis outputs.

**Passage and boundary map**

The passage contains 69 words and therefore 68 possible interword positions. Each position was classified from the punctuation attached to the canonical token immediately before it. A position was sentence-final if that token ended with a period, punctuation-internal if it ended with a comma, colon, or semicolon, and unpunctuated otherwise. The primary placement contrast combined the three sentence-final and four punctuation-internal positions into seven punctuation-marked boundaries. The other 61 positions contained no punctuation mark. Because every speaker read the same word sequence, these denominators were constant across participants. Punctuation was used as a reproducible indicator of visible textual boundary marking, not as a complete syntactic or prosodic analysis.

### Table 2. Boundary map

| Boundary category | Positions | Definition |
|---|---|---|
| Sentence-final | 3 | Previous token ends with a period |
| Punctuation-internal | 4 | Previous token ends with a comma, colon, or semicolon |
| Punctuation-marked | 7 | Sentence-final plus punctuation-internal positions |

| Boundary category | Positions | Definition |
|---|---|---|
| Unpunctuated | 61 | All other interword positions |

*Note.* The 68 positions were additionally coded by syntactic juncture; the full annotation is in S1 Appendix.

To refine the broad unpunctuated category, a rule-based annotation classified all 68 positions as within-phrase, phrase-boundary, or punctuation, following the rationale of syntax-sensitive pause-location measures [16]. Within-phrase positions included determiner–noun, adjective–noun, compound, infinitival marker–verb, and preposition–object junctures. Phrase-boundary positions separated larger constituents without an overt punctuation mark. The resulting map contained 35 within-phrase positions, 26 unpunctuated phrase boundaries, and seven punctuation-marked positions. One coder completed the annotation before the geography-corrected reanalysis. It was used as a coarse exploratory refinement and should not be treated as a multi-coder syntactic gold standard or a prosodic annotation.

**Forced alignment and gap extraction**

Each audio file was paired with the canonical passage transcript and aligned using Montreal Forced Aligner version 3.3.9 [27], the english_us_arpa dictionary, and the english_us_arpa acoustic model. The procedure produced a word-level TextGrid for every recording. For each adjacent word pair, the interword interval was calculated as the aligned start time of the later word minus the aligned end time of the earlier word. Positive intervals were retained as acoustic gaps. Zero-duration positions were recorded in alignment quality-control output, and the final alignments contained no negative intervals.

The term acoustic gap is used deliberately. A positive interval between forced-aligned word boundaries is an observable timing interval, but the interval is not automatically a manually verified silent pause: it may contain breath, low-level signal, or boundary-placement error. Gaps lasting at least 250 ms were counted in the primary analysis; gaps lasting at least 500 ms were counted in the stricter threshold analysis. Counted gaps were assigned to the fixed boundary map by word position. Cumulative duration was the sum of counted gap durations within the relevant category for each speaker.

**Statistical analysis**

Speaker-level outcomes were the count and cumulative duration of all counted gaps, punctuation-marked gaps, and unpunctuated-position gaps. Counts were modeled per completed passage rather than per minute because all speakers had the same 68 boundary opportunities and recording duration partly reflects the timing behavior under study. Each count outcome was first fit with Poisson regression. A negative binomial model was substituted when the Pearson dispersion statistic from the Poisson model exceeded 1.5, a threshold fixed in the analysis code before the final reanalysis that errs toward the more conservative family when meaningful overdispersion is present; the diagnostic and selected family for every model are reported in S1 Appendix. These count models summarize speaker-level totals and support the position-level analysis; they do not themselves encode the bounded opportunity structure, which the crossed-effects and opportunity models handle directly. Cumulative durations were right-skewed and included many zeros (39 of 58 English and 15 of 57 Mandarin speakers accumulated no unpunctuated gap duration), so they were transformed as log(1 + cumulative seconds) and analyzed with ordinary least squares using HC3 heteroskedasticity-robust standard errors, reported as supporting analyses. A two-part sensitivity analysis supplemented these models: a logistic model for producing at least one unpunctuated-position gap and a log-scale model for cumulative duration among speakers with one or more gaps (S1 Appendix).

All speaker-level models included group, age in years, and sex, with American English and female as the reference categories. The focal estimate was the Mandarin-versus-English coefficient. Count coefficients are presented with exponentiated rate ratios; duration coefficients remain on the log(1 + seconds) scale. The models estimate adjusted associations within this sample. They do not eliminate unmeasured differences in proficiency, reading experience, education, or recording context.

The primary inferential analysis was a position-level crossed-effects logistic model. Each speaker contributed one binary observation for each of the 68 interword positions (7,820 observations in total), indicating whether a counted gap occurred at that position. Fixed effects were group, boundary type, their interaction, z-standardized age, and sex; crossed random intercepts for speakers and for passage positions accounted for speaker-level differences in overall gap propensity and for position-level heterogeneity in how strongly particular word boundaries attract gaps. The models were estimated by maximum likelihood with a Laplace approximation (lme4) at both thresholds. The focal test, defined before the geography-corrected reanalysis, was the group-by-unpunctuated interaction at 250 ms; the 500 ms models and all exclusion, acoustic-confirmation, duration, and syntactic analyses were sensitivity or exploratory analyses. All tests were two-sided with $\alpha = .05$, and no multiplicity adjustment was applied because the design specifies a single focal contrast whose robustness—rather than separate hypotheses—the remaining analyses probe. As a transparent aggregated check on the same contrast, a denominator-sensitive binomial opportunity model represented each speaker twice: once for the seven punctuation-marked opportunities and once for the 61 unpunctuated opportunities. The response was the number of counted gaps out of the available positions. Predictors were group, boundary type, their interaction, z-standardized age, and sex; standard errors were clustered by speaker to account for the two rows contributed by each participant. The group-by-unpunctuated interaction tested whether the Mandarin–English difference was larger at unpunctuated positions than at punctuation-marked boundaries. This aggregated model complements the position-level analysis and is reported as a secondary check.

For the exploratory syntactic refinement, each speaker's 250 ms gaps at the 35 within-phrase positions were counted. The groups were compared with a Wilcoxon rank-sum test because the English distribution was concentrated at zero; the unadjusted ratio of group means is supplied only as a descriptive effect-size aid. All tests were two-sided. Exact p values are reported when $p \geq 0.001$, and focal estimates are accompanied by 95% confidence intervals. Analyses were conducted in R 4.4.3 with the MASS and lme4 packages. The repository scripts reproduce the crossed-effects, speaker-level, opportunity, sensitivity, acoustic-confirmation, and syntactic analyses from the derived files.

### Quality control and sensitivity analyses

Alignment quality control checked the expected 69-word count, the normalized canonical word sequence, missing or unexpected word tiers, and zero or negative interword intervals. Per-position summaries tested whether the result was driven by one word pair. Threshold sensitivity repeated the analysis at 500 ms. A nonadjacent analysis removed unpunctuated positions immediately adjacent to punctuation-marked boundaries, reducing the possibility that an interval displaced from a visible mark was counted as a qualitatively separate unpunctuated event.

Manual inspection in Praat [28] covered a purposive 10-interval quality-control set spanning groups, punctuation-marked and unpunctuated positions, and both typical and flagged cases. One inspected interval belonged to mandarin11, who was subsequently excluded by the geographic criterion; the final-sample manual summary therefore contains nine intervals. The inspection distinguished whether an audible acoustic gap was present from whether the aligner assigned the surrounding word boundaries precisely.

A corpus-wide acoustic analysis implemented with Parselmouth [29] provided a signal-based characterization of every positive aligner-defined interval. Intensity was computed with Praat's algorithm (minimum pitch 100 Hz), and a frame was classified as silent when its intensity fell more than 25 dB below the recording's 99th-percentile maximum. The criterion is independent of the aligner's implicit classification of the interval, although the interval boundaries themselves derive from the alignment. For each speaker, the mean intensity of the gap was compared with the median intensity of that speaker's speech. A counted gap was conservatively confirmed as silence when its mean intensity was at least 10 dB below the speaker-specific speech level and at least 50% of the interval was classified as silent. The main unpunctuated count model and the binomial opportunity model were then refit using confirmed gaps only.

Whisper large-v3 [30] was used only as an auxiliary transcript-deviation screen, not as a replacement transcript or ground truth. Recordings were processed with and without voice-activity detection. The union of cases classified as severe under either setting comprised english361, mandarin24, english204, and english95. Excluding these four records after the geographic correction left 111 speakers. This conservative sensitivity analysis tested whether departures from the canonical reading could account for the main placement contrast. Together, the checks address different failure modes: threshold dependence, local concentration, positional misalignment, intervals that are positive in the TextGrid but not acoustically silence-like, and transcript deviation.

### Use of artificial intelligence tools

Anthropic Claude and OpenAI Codex were used during manuscript preparation in 2026 to assist with language editing, prose organization, formatting, drafting and organizing analysis and verification code, consistency checks, and construction of submission and quality-assurance files. Consensus was used to locate candidate scholarly literature; every retained source was read and verified against the original publication. AI tools assisted in organizing and cross-checking participant metadata, but the author personally verified every archive record against its source page and made all eligibility decisions. The author evaluated AI-assisted outputs by checking every retained citation against publisher or bibliographic records, reviewing all code, reconciling reported values to the source data and R outputs, rerunning the complete analysis, and visually inspecting all generated documents. AI-assisted changes affected wording, organization, formatting, code, and verification materials; they did not generate or alter the original recordings or derived measurements, define eligibility rules or outcomes, select statistical conclusions, or replace author interpretation. The author reviewed and approved all content and accepts full responsibility for the article.

## Results

### Alignment and sample quality

The final sample contained 58 American English and 57 Mandarin speakers. Mean age was 35.65 years in the English group and 29.16 years in the Mandarin group; the sex counts were 32 female and 26 male speakers in English and 35 female and 22 male speakers in Mandarin (Table 1). Recordings averaged 22.97 s and 31.52 s, respectively. These duration and nominal-rate differences describe the complete readings and were not interpreted as articulation-rate estimates.

All 115 analyzed TextGrids matched the expected 69-word count and normalized word sequence. No retained recording had a missing or unexpected word tier, and no negative interword gap was detected. With voice-activity detection enabled, the auxiliary transcript screen classified 99 recordings as clean, 14 as minor or moderate deviations, and two as severe mismatches. Without voice-activity detection, the corresponding counts were 87, 25, and three. Audio-quality checks identified no very short, very long, or clipping-flagged recording.

#### Descriptive placement results

At the 250 ms threshold, Mandarin speakers produced 9.23 counted gaps per passage on average and English speakers produced 5.90 (Table 3). The difference was not distributed proportionally across the boundary map. Punctuation-marked means were 5.56 and 5.47 for the Mandarin and English groups, respectively, whereas unpunctuated-position means were 3.67 and 0.43. The distributions also differed in their lower tails: the median unpunctuated count was 2 (interquartile range 0–6) for Mandarin speakers and 0 (interquartile range 0–1) for English speakers. Mandarin speakers accumulated 1.91 s of unpunctuated gap duration on average, compared with 0.18 s for English speakers. Fig 1 displays the placement pattern across all 68 interword positions and the speaker-level distributions.

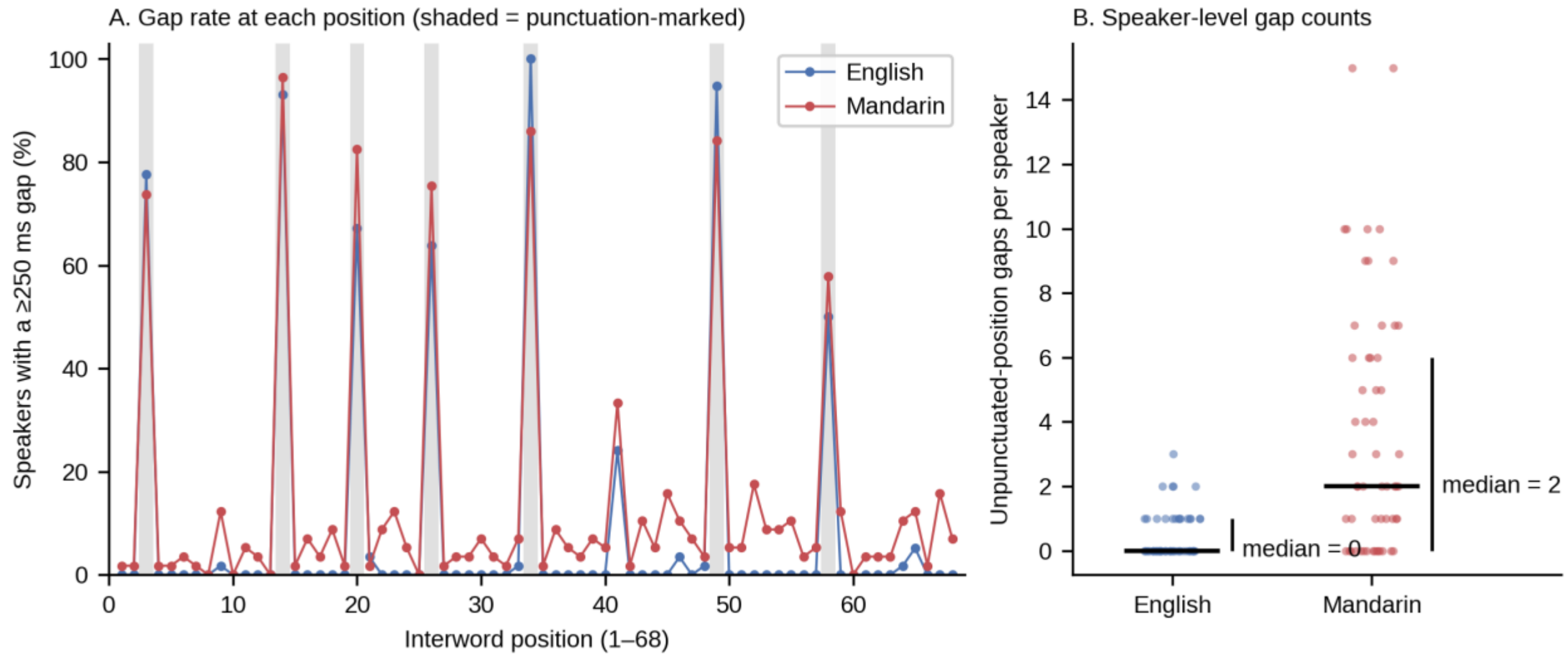

**Fig 1. Acoustic gap placement in the fixed passage.** (A) Percentage of speakers in each group producing a counted (≥250 ms) gap at each of the 68 interword positions; shaded bands mark the seven punctuation-marked positions. (B) Speaker-level counts of unpunctuated-position gaps, with group medians (horizontal lines) and interquartile ranges (vertical bars).

**Table 3. Descriptive acoustic-gap placement**

| Threshold | Group | n | Total count, mean (SD) | Punctuation count, mean (SD) | Unpunctuated count, mean (SD) | Unpunctuated cumulative duration, mean (SD), s |
|---|---|---|---|---|---|---|
| 250 ms | American English L1 | 58 | 5.90 (1.79) | 5.47 (1.42) | 0.43 (0.70) | 0.18 (0.33) |
| 250 ms | Mandarin L1 | 57 | 9.23 (3.68) | 5.56 (1.41) | 3.67 (3.91) | 1.91 (2.30) |
| 500 ms | American English L1 | 58 | 2.64 (2.05) | 2.55 (1.96) | 0.09 (0.34) | 0.06 (0.23) |
| 500 ms | Mandarin L1 | 57 | 4.30 (3.02) | 2.75 (1.71) | 1.54 (2.34) | 1.16 (1.85) |

*Note.* Values are speaker-level means with standard deviations in parentheses.

At the stricter 500 ms threshold, the same descriptive asymmetry remained. Punctuation-marked means were 2.75 for Mandarin and 2.55 for English speakers, while unpunctuated means were 1.54 and 0.09. Expressed against the 61 available unpunctuated positions, the mean 250 ms opportunity rates were 6.01% for Mandarin and 0.71% for English speakers; at 500 ms they were 2.53% and 0.14%. These percentages summarize speaker-level opportunity use and are not separate inferential tests.

**Primary models**

The primary crossed-effects model located the group difference at unpunctuated positions while accounting for both speaker and position heterogeneity. At 250 ms, the group-by-unpunctuated interaction was b = 2.25 (SE = 0.30), odds ratio 9.44, 95% CI 5.23–17.02, $p < 0.001$, with random-intercept standard deviations of 0.74 for speakers and 0.87 for positions; the Mandarin main effect at punctuation-marked positions was not significant (odds ratio 1.31, 95% CI 0.82–2.08, $p = 0.262$). At 500 ms, the interaction was b = 3.01 (SE = 0.51), odds ratio 20.30, 95% CI 7.44–55.41, $p < 0.001$. Because the position random intercepts adjust for baseline differences in gap propensity across word boundaries, the interaction is not explained merely by certain positions attracting gaps frequently in both groups (full estimates in S1 Appendix).

Age- and sex-adjusted count models estimated a Mandarin-versus-English rate ratio of 1.67 for all gaps (95% CI 1.45–1.93), 1.03 for punctuation-marked gaps (95% CI 0.88–1.21), and 11.98 for unpunctuated-position gaps (95% CI 6.53–22.00; Table 4). Thus, the overall count difference combined a near-null punctuation-marked contrast with a large unpunctuated contrast. The corresponding punctuation-marked coefficient was small and nonsignificant ($p = 0.724$), whereas the unpunctuated coefficient remained clearly positive despite the wider negative-binomial confidence interval.

**Table 4. Primary 250 ms group coefficients**

| Outcome | Model | b | SE | 95% CI | p | Rate ratio |
|---|---|---|---|---|---|---|
| Total gap count | Poisson | 0.52 | 0.07 | 0.37 to 0.66 | <0.001 | 1.67 |
| Punctuation-marked count | Poisson | 0.03 | 0.08 | −0.13 to 0.19 | 0.724 | 1.03 |
| Unpunctuated-position count | Negative binomial | 2.48 | 0.31 | 1.88 to 3.09 | <0.001 | 11.98 |
| log(1 + total duration) | OLS HC3 | 0.40 | 0.07 | 0.26 to 0.54 | <0.001 | — |
| log(1 + punctuation duration) | OLS HC3 | 0.06 | 0.06 | −0.05 to 0.17 | 0.286 | — |
| log(1 + unpunctuated duration) | OLS HC3 | 0.74 | 0.10 | 0.54 to 0.93 | <0.001 | — |

*Note.* Coefficients compare Mandarin with American English and control for age and sex. Rate ratios are exp(b) for count models. Full coefficients and model-selection diagnostics are in S1 Appendix.

Cumulative-duration models showed the same placement-specific pattern. The adjusted Mandarin coefficient was 0.74 for log(1 + unpunctuated seconds), 95% CI 0.54–0.93, but only 0.06 for log(1 + punctuation-marked seconds), 95% CI −0.05–0.17. Accordingly, the placement result was not confined to the number of thresholded events; it was also present in the accumulated duration of those events. The two-part sensitivity analysis agreed: the odds of producing any unpunctuated-position gap were higher in the Mandarin group (odds ratio 9.24, 95% CI 3.61–23.61), cumulative duration among speakers with at least one such gap was longer (log-scale $b = 1.51$, 95% CI 1.08–1.94), and a rank-based comparison of the full duration distributions led to the same conclusion (Mann–Whitney $p < 0.001$; S1 Appendix).

The aggregated opportunity model agreed with the crossed-effects analysis. At 250 ms, the group-by-unpunctuated interaction was $b = 2.25$ (clustered $SE = 0.35$), 95% CI 1.56–2.95, $p < 0.001$, corresponding to an odds ratio of 9.50 (95% CI 4.74–19.03). The Mandarin main effect at punctuation-marked positions was not significant (odds ratio 1.25, 95% CI 0.78–2.02, $p = 0.355$). At 500 ms, the interaction remained positive, $b = 2.83$ ($SE = 0.56$), 95% CI 1.73–3.92, $p < 0.001$, odds ratio 16.86 (95% CI 5.63–50.51), while the punctuation-marked main effect again was not significant. The inference from the fixed opportunity denominators therefore agreed with the separate count models.

#### Robustness and acoustic validation

Unpunctuated gaps occurred at 56 distinct passage positions at 250 ms and 37 positions at 500 ms; the pattern was therefore not produced by a single repeated word pair. Removing every unpunctuated position adjacent to punctuation reduced the available set but left the contrast largely unchanged. At 250 ms, nonadjacent means were 3.21 for Mandarin and 0.36 for English speakers, and the adjusted count coefficient was $b = 2.54$, 95% CI 1.91–3.16, $p < 0.001$. At 500 ms, the corresponding means were 1.25 and 0.09 and the coefficient was $b = 3.77$, 95% CI 2.36–5.18, $p < 0.001$.

Excluding the four severe-union transcript cases retained 111 speakers and strengthened the primary contrast: the 250 ms unpunctuated count coefficient was $b = 2.66$, 95% CI 2.02–3.29, $p <$ 0.001, while the punctuation-marked coefficient remained near zero ($b = 0.03$, $p = 0.701$). The nine manually inspected intervals from final-sample speakers were all audible acoustic gaps; six had clear intended word-pair mapping and three had real gaps with imperfect TextGrid word-boundary placement.

The geography-corrected sample contained 868 counted gaps at 250 ms: 634 punctuation-marked and 234 unpunctuated. Median intensity drops below each speaker's speech level were 23.3 dB and 22.5 dB, and median proportions of the interval classified as silent were 1.00 and 0.96, respectively. The conservative acoustic criterion confirmed 568/634 punctuation-marked gaps (89.6%) and 195/234 unpunctuated gaps (83.3%). Confirmation rates for unpunctuated gaps were 88.0% in English and 82.8% in Mandarin recordings. Restricting the analysis to confirmed silence preserved the unpunctuated count effect (rate ratio 11.30, 95% CI 5.78–22.11) and strengthened the group-by-unpunctuated opportunity interaction (odds ratio 13.96, 95% CI 7.02–27.73). The placement asymmetry therefore did not depend on intervals that failed the independent silence criterion.

The absence of a punctuation-marked group difference was not a simple ceiling artifact. English speakers paused at 5.47 of seven punctuation-marked positions on average and Mandarin speakers at 5.56; 19/58 and 20/57 speakers, respectively, paused at all seven.

#### Syntactic placement

Of the 209 Mandarin-group unpunctuated gaps at 250 ms, 101 (48.3%) occurred at positions classified as within-phrase; the remaining 108 occurred at unpunctuated phrase-boundary positions. Within-phrase means were 1.77 gaps per Mandarin speaker and 0.07 per English speaker (Wilcoxon rank-sum $p < 0.001$; unadjusted mean ratio 25.7). Thirty-two of 57 Mandarin speakers and four of 58 English speakers produced at least one within-phrase gap, and the English group produced only four such gaps in total. This result shows that the broad unpunctuated contrast was not restricted to positions annotated as boundaries between larger constituents. Its magnitude should nevertheless be read in light of the English floor and the single-coder, rule-based annotation.

## Discussion

The central result is a placement asymmetry. The Mandarin group did not simply show a proportional increase in gaps throughout the passage. Punctuation-marked counts and durations were similar to those of the English group after age and sex adjustment, whereas the additional gaps were concentrated at unpunctuated positions. This conclusion was supported by four complementary analyses: a position-level crossed-effects model with speaker and position random intercepts, separate count and duration models, a denominator-sensitive opportunity model, and models restricted to acoustically confirmed silence. Because passage positions were modeled as a crossed random factor, the interaction reflects a group difference estimated after accounting for baseline heterogeneity in gap propensity across positions, rather than a property of a few gap-attracting word pairs. The convergence matters because each analysis addresses a different alternative explanation. The result is not merely that one group had a higher passage-level total; it identifies the portion of the fixed boundary map where that difference arose.

This distinction refines the measurement of breakdown fluency. Standard totals remain useful summaries of how often or how long speech is interrupted, but they are many-to-one mappings: different spatial distributions across a text can produce the same total. Placement-sensitive measures preserve information that totals discard. This emphasis on interpretability parallels recent language-assessment work arguing that measurement is more useful when aggregate outcomes are linked to theoretically specified linguistic or cognitive features rather than treated as opaque totals [31]. In the present data, the total-count rate ratio of 1.67 was the aggregate of a punctuation-marked rate ratio near one and an unpunctuated rate ratio of 11.98. Reporting only the total would therefore understate the structural specificity of the group difference. The duration models led to the same conclusion, indicating that the result was not an artifact of thresholded counts alone.

The pattern is consistent with the broader pause-location literature without simply reproducing it. Studies of learner speech have distinguished pauses within clauses or utterances from those at larger boundaries [11,12,15], and listeners respond differently to pauses depending on their location [13]. Coulange and de Jong's syntactic-distance measure further demonstrates that location can be represented more informatively than by a single global frequency [16]. The current study contributes a simpler, fully auditable contrast suited to a fixed written passage: visible punctuation versus the complete set of unpunctuated interword positions. It does not replace continuous syntactic measures, but it establishes the main placement asymmetry before applying the coarser within-phrase refinement.

The comparison with Alfaifi and colleagues is especially informative [14]. Their smaller Mandarin–English read-speech study began from selected punctuation and prosodic locations and related pause duration to syntactic and lexical characteristics. The present extraction begins from all interword positions. That broader denominator revealed that the strongest group contrast in this corpus was not at the visible punctuation sites but away from them. This is an extension in analytic coverage rather than a contradiction: selected-boundary and full-passage analyses ask different questions and can capture different parts of the same timing distribution.

Read speech is central to what can be inferred. Holding the word sequence and punctuation constant makes boundary-level comparison possible and removes between-speaker differences in lexical choice and text structure. The strong effect of punctuation on pausing in prior read-speech research [18] was also evident here: both groups used most punctuation-marked opportunities, and their mean counts were almost identical. At the same time, neither group was at an absolute ceiling; many speakers left at least one punctuation opportunity unused. The near-null punctuation contrast is therefore more plausibly interpreted as a shared feature of these readings than as a mathematical consequence of the seven-position bound.

The methodological contribution is the validation chain. Forced alignment enables passage-wide extraction at a scale that would be costly to annotate manually [27], but an aligned positive interval cannot be assumed to be a silent pause. The manual check confirmed that all inspected final-sample intervals contained audible gaps while also showing that the surrounding word boundary could be imperfectly placed. The independent acoustic analysis then evaluated every extracted positive interval and confirmed most counted events under a conservative intensity-and-silence rule. Finally, the confirmed-only models reproduced the focal contrast. The appropriate interpretation is therefore neither that the aligner provided manual phonetic ground truth nor that its output was accepted without verification. The study measures forced-alignment-based acoustic gaps whose aggregate result survives an acoustic criterion that is independent of the aligner's silence classification, applied to intervals that the alignment itself defines. This emphasis on verification has a broader parallel in research on technology-supported problem solving, where the usefulness of computational assistance depends on critical evaluation and informed human judgment [32].

The syntactic refinement adds a second form of placement evidence. Nearly half of the Mandarin group's unpunctuated gaps occurred at positions the rule-based annotation classified as within-phrase. This result is compatible with the rationale for location- and syntax-sensitive fluency measures [15,16], but two features require restraint. First, the English distribution was close to a floor, with only four within-phrase gaps. The large mean ratio therefore describes a real distributional imbalance in this passage but is statistically unstable as a general comparative quantity. Second, one coder assigned the categories, and the binary within-phrase/phrase-boundary distinction does not represent graded syntactic distance or prosodic constituency. The analysis shows that visible punctuation and major unmarked boundaries do not exhaust the observed distribution; it does not establish a general syntactic-processing difference between populations.

Several explanations for the group association remain open. Speakers may differ in English proficiency, reading skill, educational experience, familiarity with read-aloud tasks, language history, local recording conditions, or other characteristics not standardized by the archive. The longer average recordings in the Mandarin group are part of the observed timing profile, but duration alone cannot determine why particular gaps occurred. Age and sex adjustment addresses only two measured covariates and should not be mistaken for control of proficiency or language experience. No experimental manipulation, Mandarin-language baseline, or within-speaker comparison was available to identify a causal mechanism. More broadly, interpreting language performance requires attention to the educational opportunities and assessment contexts through which competence is developed and recognized [33].

Sampling also limits generalization. The analysis is based on a purposive project corpus, not a probability sample. The sample was exhaustive under the stated eligibility rules at the corpus-assembly date, but it is not a current census of the continually updated archive. The geographic eligibility audit made the intended rules explicit and removed three non-mainland records independently of their outcomes, but it does not convert the sample into a population census. The findings characterize these 115 public recordings and should not be interpreted as prevalence estimates for Mandarin speakers, mainland-China-born learners of English, or Midwest English speakers as populations.

Other limitations concern task and validation scope. One 69-word passage supplies only 68 repeated positions, and its punctuation, lexical sequence, and syntactic structures are idiosyncratic. Punctuation is an objective visible cue but an incomplete proxy for syntax and prosody. The manual validation set was small, and full manual boundary annotation was not feasible. The transcript screen was auxiliary and does not prove canonical production word by word. Finally, the study collected no listener judgments. It therefore cannot establish effects on perceived fluency, comprehensibility, or intelligibility, which are listener-based outcomes distinct from the timing measures analyzed here [34,35]. This distinction reflects a broader concern in educational analytics: observable behavioral indicators require independent validation before they can support conclusions about the underlying abilities or difficulties they are intended to represent [36]. Complementary evidence can be obtained by combining performance measures with qualitative accounts of learners’ conceptual understanding, as illustrated by mixed methods research in computer science education [37]. More generally, short-term changes in self-reported behavior require independent assessment and follow-up before they can support broader conclusions about intervention outcomes [38].

Future research should test the generality and usefulness of placement measures rather than assume them. Stronger designs would preregister dated archive-sampling rules; include multiple read passages and spontaneous tasks; collect standardized proficiency, literacy, and language-history measures; and obtain larger manual annotations. Multi-coder or dependency-based syntax measures could replace the present binary refinement. Listener studies could then determine whether full-passage placement indices explain perceived fluency or comprehensibility beyond global count, duration, and speed measures. A within-speaker design across languages or tasks would further help separate stable speaking style from task- and second-language-specific timing. Placement-sensitive measures could also complement evaluations of reading-based instruction, extending research on learning and retention [39] to examine whether instructional gains are accompanied by changes in the temporal organization of speech. More broadly, examining learners' engagement during language tasks can complement evaluations of their resulting performance, as illustrated by research on interactional patterns and feedback focus in classroom activities [40].

## Conclusion

In this fixed-passage archival corpus, the Mandarin and American English groups differed primarily in the frequency and cumulative duration of interword acoustic gaps at unpunctuated positions, not at punctuation-marked boundaries. The placement contrast remained under a stricter duration threshold, after removing punctuation-adjacent positions and severe transcript-deviation cases, and when inference was restricted to acoustically confirmed silence. Nearly half of the Mandarin group's unpunctuated gaps occurred at positions classified as within-phrase, although that exploratory result is limited by a single coder and an English floor. Taken together, the findings support acoustic gap placement as a reproducible, observable dimension of breakdown fluency in fixed-passage read speech. They do not identify the cause of the group association or establish a proficiency or perceptual consequence.

## Data availability

Original recordings and public metadata are available from the Speech Accent Archive (https://accent.gmu.edu/). The article does not redistribute audio. All derived participant-level and gap-level data, boundary annotations, model inputs, complete model outputs, and analysis code needed to reproduce the reported findings are provided as S2 Data and S3 File. If the article is accepted, the S3 reproducibility package will additionally be deposited in the Open Science Framework, and its permanent DOI and URL will be added before publication.

## Acknowledgments

The author thanks the Speech Accent Archive and its contributors for making the recordings available for research.

## Supporting information captions

**S1 Appendix.** Supporting methods, quality-control results, and full statistical tables.

**S2 Data.** Participant metadata, selection audit, and source URLs for the final analytic sample.

**S3 File.** Reproducibility package containing derived gap-level data, boundary annotations, model inputs, R code, complete model outputs, a data dictionary, and a README.